\documentclass[twocolumn,showpacs,preprintnumbers,amsmath,amssymb]{revtex4}
\usepackage{amssymb}

\usepackage{graphicx}
\usepackage{dcolumn}
\usepackage{bm}

\begin{document}

\title{Experimental investigation of classical and quantum correlations under decoherence}
 \author{Jin-Shi Xu}
 \affiliation{Key Laboratory of Quantum Information,
  University of Science and Technology
  of China, CAS, Hefei, 230026, People's Republic of China}

\author{Xiao-Ye Xu}
\affiliation{Key Laboratory of Quantum Information, University of
Science and Technology of China, CAS, Hefei, 230026, People's
Republic of China}

\author{Chuan-Feng Li$\footnote{email: cfli@ustc.edu.cn}$}
\affiliation{Key Laboratory of Quantum Information,
  University of Science and Technology
  of China, CAS, Hefei, 230026, People's Republic of China}

\author{Cheng-Jie Zhang}
\affiliation{Key Laboratory of Quantum Information, University of
Science and Technology of China, CAS, Hefei, 230026, People's
Republic of China}

\author{Xu-Bo Zou$\footnote{email: xbz@ustc.edu.cn}$}
\affiliation{Key Laboratory of Quantum Information,
  University of Science and Technology
  of China, CAS, Hefei, 230026, People's Republic of China}

 \author{Guang-Can Guo}
\affiliation{Key Laboratory
of Quantum Information, University of Science and Technology of
China, CAS, Hefei, 230026, People's Republic of China}
\date{\today }

\begin{abstract}
It is well known that many operations in quantum information
processing depend largely on a special kind of quantum correlation,
that is, entanglement. However, there are also quantum tasks that
display the quantum advantage without entanglement. Distinguishing
classical and quantum correlations in quantum systems is therefore
of both fundamental and practical importance. In consideration of
the unavoidable interaction between correlated systems and the
environment, understanding the dynamics of correlations would
stimulate great interest. In this study, we investigate the dynamics
of different kinds of bipartite correlations in an all-optical
experimental setup. The sudden change in behaviour in the decay
rates of correlations and their immunity against certain
decoherences are shown. Moreover, quantum correlation is observed to
be larger than classical correlation, which disproves the early
conjecture that classical correlation is always greater than quantum
correlation. Our observations may be important for quantum
information processing.
\end{abstract}

\pacs{03.67.Yz, 42.50.Xa, 03.65.Ud}
\maketitle

\section{Introduction}
Correlations, including classical and quantum parts, are crucial in
science and technology. Although many operations in quantum
information processing largely depend on a special kind of quantum
correlation, that is, entanglement \cite{Nielsen00}, there are
quantum tasks that display the quantum advantage without
entanglement \cite{Meyer00,Kenigsberg06,Datta08}, and some of these
have been verified experimentally \cite{Lanyon08}. Distinguishing
classical and quantum correlations is therefore of both fundamental
and practical importance. Many theoretical studies have been
conducted in this direction
\cite{Henderson01,Ollivier01,Oppenheim02,Terhal02,DiVincenzo04,Horodecki05,Groisman05,Luo08,Modi10}.
Among them, quantifying the quantumness of correlations with quantum
discord \cite{Ollivier01} has received great attention
\cite{Datta08,Lanyon08,Zurek03,Rodriguez08,Piani08,Luo082,Werlang09,Ferraro09,Wang10,Fanchini09,Maziero09,Maziero092,Sarandy09,Luo09}.

In the field of quantum information, for a bipartite system
$\rho_{AB}$, it is widely accepted that quantum mutual information
measures its total correlations \cite{Groisman05,Schumacher06}
defined as
$\mathcal{I}(\rho_{AB})=S(\rho_{A})+S(\rho_{B})-S(\rho_{AB})$
\cite{Vedral02}, where $\rho_{A}$ and $\rho_{B}$ are the
reduced-density matrices of $\rho_{AB}$.
$S(\rho)=-\operatorname*{tr}(\rho\log_{2}\rho)$ in the von Neumann
entropy. Depending on the maximal information gained for $\rho_{AB}$
with measurement on one of the subsystems, classical correlation
($\mathcal{C}$) \cite{Henderson01} is defined as
$\mathcal{C}(\rho_{AB})\equiv\underset{B_{j}^{\dag}B_{j}}{\max}[
S(\rho_{A})-\underset{j}\sum q_{j} S(\rho_{A}^{j})]$ , where
$B_{j}^{\dag}B_{j}$ is a positive-operator-valued measure performed
on the subsystem $B$ and
$q_{j}=\operatorname*{tr}_{AB}(B_{j}\rho_{AB}B_{j}^{\dag})$.
$\rho_{A}^{j}=\operatorname*{tr}_{B}(B_{j}\rho_{AB}B_{j}^{\dag})/q_{j}$
is the postmeasurement state of $A$ after obtaining the outcome $j$
on the particle $B$. Quantum correlation is therefore given by
$\mathcal{Q}(\rho_{AB})=\mathcal{I}(\rho_{AB})-\mathcal{C}(\rho_{AB})$.
In such a case, $\mathcal{Q}$ is just identical to quantum discord
with a definition based on the distinction between classical
information theory and quantum information theory \cite{Ollivier01},
which can be further distributed into entanglement and the
additional quantum correlation (non-entanglement quantum
correlation) \cite{Vedral092}.

Because of the unavoidable interaction between a quantum system and
its environment, understanding the dynamics of different kinds of
correlations has stimulated great interest. Some studies have
focused on the comparison between the dynamics of quantum discord
and entanglement under both Markovian \cite{Werlang09} and
Non-Markovian \cite{Wang10,Fanchini09} environments. Their
behaviours have been shown to be very different, and the quantum
discord is more robust against decoherence than entanglement in the
Markovian evolution \cite{Werlang09}. Recently, several peculiar
properties in the dynamics of classical and quantum correlations
have also been shown with the present of Markovian noise
\cite{Maziero09,Maziero092}, in which the decay rates of
correlations may exhibit sudden changes in behaviour
\cite{Maziero09} and the bipartite quantum correlation may
completely disappear without being transferred to the environment
\cite{Maziero092}.

In this study, we experimentally investigate the dynamics of
classical and quantum correlations between biqubit systems in a
one-sided phase-damping channel. The sudden changes in behaviours in
the decay rates of classical and quantum correlations are shown in
an all-optical experimental setup, in which the dephasing
environment is simulated by birefringent quartz plates. Classical
and quantum correlations are observed to remain unaffected under
certain decoherence areas. Moreover, quantum correlation is shown to
be larger than classical correlation during the dynamics of a
special input state, which contradicts the early conjecture that
classical correlation is always greater than quantum correlation
\cite{Lindblad91}. Our observations may have important roles in
quantum information processing.

\section{Results}
{\bf Theoretical schemes.} Consider a two-level quantum system $S$
with lower and upper states $|0\rangle_{S}$ and $|1\rangle_{S}$
under the action of a phase-damping environment $E$ ($|0\rangle_{E}$
is the initial state). The interaction quantum map can be written as
\cite{Preskill98}
\begin{eqnarray}
|0\rangle_{S}\otimes|0\rangle_{E}&\rightarrow&\sqrt{1-p}|0\rangle_{S}\otimes|0\rangle_{E}+\sqrt{p}|0\rangle_{S}\otimes|1\rangle_{E},\nonumber \\
|1\rangle_{S}\otimes|0\rangle_{E}&\rightarrow&\sqrt{1-p}|1\rangle_{S}\otimes|0\rangle_{E}+\sqrt{p}|1\rangle_{S}\otimes|2\rangle_{E}.
\label{quantum map}
\end{eqnarray}
Under such a uniquely quantum mechanical noise process, the system
$S$ remains in the initial state with the probability $p$ of
scattering the environment to its excited state $|1\rangle_{E}$ if
$S$ is in state $|0\rangle_{S}$ and to $|2\rangle_{E}$ if $S$ is in
state $|1\rangle_{S}$. It physically describes, for instance, the
case when a photon scatters randomly in a fibre. The coherence of
$S$ degrades exponentially, and $p=1-\exp(-\Gamma t)$, where
$\Gamma$ is the decay rate. When it extends to the case of bipartite
systems coupling with this environment, the decoherence-free states
\cite{Kwiat00} and the phenomenon of entanglement collapse and
revival \cite{Xu09} have been demonstrated.

When an initial state
$\rho_{AB}=a|\Phi^+\rangle\langle\Phi^+|+b|\Phi^-\rangle\langle\Phi^-|
+c|\Psi^+\rangle\langle\Psi^+|+d|\Psi^-\rangle\langle\Psi^-|$, with
$|\Phi^\pm\rangle=1/\sqrt{2}(|00\rangle\pm|11\rangle)$ and
$|\Psi^\pm\rangle=1/\sqrt{2}(|01\rangle\pm|10\rangle)$ representing
the four Bell states and $a+b+c+d=1$, evolves in the phase-damping
channel given by equation (\ref{quantum map}), its off-diagonal
elements decay exponentially according to the previous analysis.
Because $S(\rho_{A})=S(\rho_{B})=1$, the total correlation is
calculated as
\begin{equation}
\mathcal{I}(\rho_{AB})=2+\sum_{j=1}^{4}\lambda_{j}\log_{2}\lambda_{j},
\label{total}
\end{equation}
where $\lambda_{j}$ is the four eigenvalues of the final density
matrix. To calculate the classical correlation \cite{Henderson01},
the state in mode $B$ is projected into
$|l\rangle=\cos\theta|0\rangle+\sin\theta e^{i\phi}|1\rangle$ to get
the minimal conditional entropy of subsystem $A$, where
$0\leq\theta\leq\pi$ and $0\leq\phi\leq2\pi$. In the biqubit case,
the projective measurement is the positive-operator-valued measure
to maximize $\mathcal{C}(\rho_{AB})$ \cite{Hamieh04}, which is
expressed as \cite{Maziero09}
\begin{align}
\mathcal{C}(\rho_{AB})&=S(\rho_{A})-\underset{\theta,\phi}{\min}[S(\rho_{A}^{l})] \nonumber \\
&=\frac{1-\eta}{2}\log_{2}(1-\eta)+\frac{1+\eta}{2}\log_{2}(1+\eta),
\label{classical}
\end{align}
$S(\rho_{A}^{l})$ represents the conditional entropy of $A$ with
projecting $B$ in $|l\rangle$ and
$\eta=\max\{|\alpha|,|\beta|,|\gamma|\}$ with
$\alpha=(1-p)(a-b+c-d)$, $\beta=(1-p)(c-d-a+b)$ and
$\gamma=a+b-c-d$. The quantum correlation is therefore given by
\begin{equation}
\mathcal{Q}(\rho_{AB})
=2+\sum_{j=1}^{4}\lambda_{j}\log_{2}\lambda_{j}-\mathcal{C}(\rho_{AB}).
\label{quantum}
\end{equation}

{\bf Experimental demonstration of the dynamics of correlations.}
Photon qubits with polarization encoded as the information carriers
have been widely used to implement different quantum information
processing procedures \cite{Nielsen00}. The coupling between photon
polarization and frequency modes in a birefringent environment leads
to dephase by a trace-over frequency \cite{Berglund00}, which
simulates the decoherence effect in equation (\ref{quantum map}).
Here, we encode the horizontal ($H$) and vertical ($V$)
polarizations of a photon as $|0\rangle_{S}$ and $|1\rangle_{S}$ of
a qubit and pass the polarization-entangled photons through
one-sided controllable quartz plates to investigate the dynamics of
different kinds of correlations.

\begin{figure}[tbph]
\begin{center}
\includegraphics [width= 3.0in]{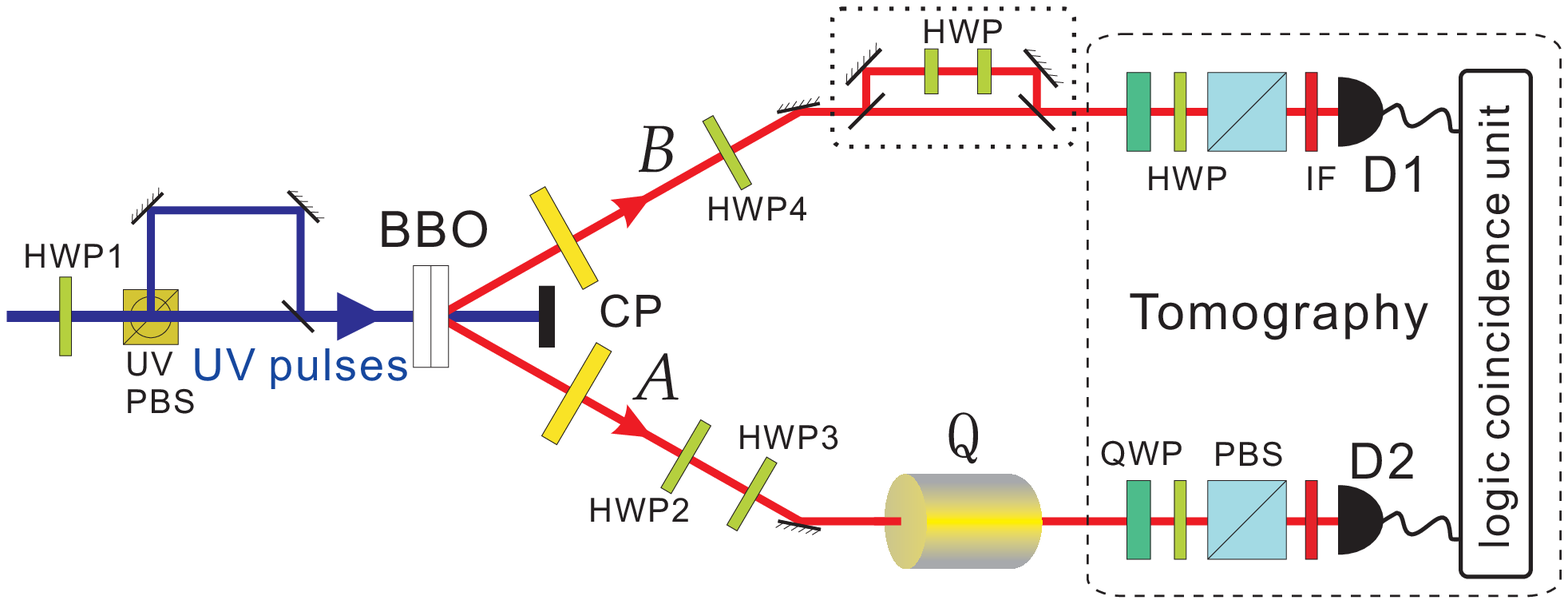}
\end{center}
\caption{(Color online). Experimental setup. Ultraviolet (UV) pulses
are divided into two parts by the ultraviolet polarization beam
splitter (UV PBS). The relative power between them can be changed by
a half-wave plate (HWP1). Entangled photon pairs are prepared by
spontaneous parametric down conversion by pumping the two adjacent
crystals ($\beta$-barium borate, BBO). Quartz plates (CP) are used
to compensate the birefringence of down conversion photons, both of
which then pass through half-wave plates (HWP2 and HWP4) with the
optic axes set at $22.5^{\circ}$. The photon in mode $A$ further
passes through a half-wave plate (HWP3) with the optic axis set as
horizontal to prepare the required mixed state. The dephasing
environment is simulated by quartz plates (Q) with thickness $L$. An
unbalanced Mach-Zehnder device in the dotted pane with the
reflective part further passing through two half-wave plates (HWP)
is inserted into mode $B$ to prepare another initial mixed state.
Quarter-wave plates (QWP), half-wave plates (HWP), and polarization
beam splitters (PBS) are used in each mode to set the detecting
bases for implementing state tomography. Single photon detectors
equipped with interference filters (IF) are used to detected these
two photons.} \label{fig:setup}
\end{figure}

Figure 1 shows our experimental setup. Ultraviolet pulses are
frequency doubled from a mode-locked Ti:sapphire laser with
wavelength centred at 780 nm, with a 130 fs pulse width and a 76 MHz
repetition rate. They are distributed into two paths by an
ultraviolet polarization beam splitter, which transmits $45^\circ$
linearly polarized photons ($1/\sqrt{2}(H+V)$) and reflects
$-45^\circ$ linearly polarized photons ($1/\sqrt{2}(H-V)$). The
relative power between these two paths can be changed easily by a
half-wave plate (HWP1), and the time difference between them is
about 6 ns. They then combine again by a beam splitter, and both
pump two identically cut type-I $\beta$-barium borate crystals, with
their optic axes aligned in mutually perpendicular planes, to
produce polarization-entangled photon pairs \cite{Kwiat99}. After
compensating the birefringence effect between $H$ and $V$ with
quartz plates (CP), the short pump beam produces the maximally
entangled state $|\Phi^{+}\rangle$, whereas the long pump beam
produces $|\Phi^{-}\rangle$.

Two HWPs (HWP2 and HWP4), with the optic axes set to $22.5^\circ$,
can change $H$ into $1/\sqrt{2}(H+V)$ and $V$ into
$1/\sqrt{2}(H-V)$. The HWP (HWP3) with the optic axis set as
horizontal in mode $A$ is used to introduce $\pi$-phase between $H$
and $V$. As our detection traces over the time difference between
these two production processes, as has been shown earlier
\cite{Puentes06,Aiello07}, the prepared state becomes
\begin{equation}
\rho_{AB}=b|\Phi^-\rangle\langle\Phi^-|+d|\Psi^-\rangle\langle\Psi^-|, \label{state-interference}%
\end{equation}
where $b$ is determined by the relative power between these two pump
beams and $b+d=1$. Then the photon in mode $A$ with frequency
distribution $f(\omega)$ passes through the dephasing channel, which
is stimulated by quartz plates (Q) with the optic axis set to be
horizontal and with a thickness of $L$. Therefore, the decoherence
parameter $\kappa=\int f(\omega) \exp(i\tau\omega) \mathrm{d}\omega$
imposes on the off-diagonal elements of equation
(\ref{state-interference}), where $\tau=L\Delta n/c$, with $c$
representing the vacuum velocity of the photon and $\Delta n$ is the
difference between the indices of refraction of $H$ and $V$. In this
case, we get that $p=1-|\kappa|$ in equation (\ref{quantum map}).

An unbalanced Mach-Zehnder device in the dotted pane, which further
separates the photon into long and short paths, is inserted into
mode $B$ to prepare another input state. The long path in the dotted
pane contains a HWP with the optic axis set at $45^\circ$ and
another HWP with the optic axis set as horizontal. The time
difference between these two paths is much larger than the coherent
time of the photon and less than the coincidence window of the logic
circuit (at about 3 ns). As a result, the prepared state becomes
\begin{align}
\rho&=dR|\Phi^+\rangle\langle\Phi^+|+b(1-R)|\Phi^-\rangle\langle\Phi^-| \nonumber \\
&+bR|\Psi^+\rangle\langle\Psi^+|+d(1-R)|\Psi^-\rangle\langle\Psi^-|,
\label{four-final}
\end{align}
where $R$ is the effective total reflectivity of the two partial
reflecting mirrors in the dotted pane. The photon in mode $A$ then
further passes through the dephasing channel. Finally, the evolved
state is reconstructed by tomography. Quarter-wave plates, HWPs and
polarization beam splitters are used in each mode to set the usual
16 measurement bases \cite{James01}. These two photons are detected
by single-photon detectors equipped with 3 nm (full width at half
maximum) interference filters.

\begin{figure}
\centering
\begin{minipage}[c]{0.5\textwidth}
\centering
\includegraphics[width=2.5in]{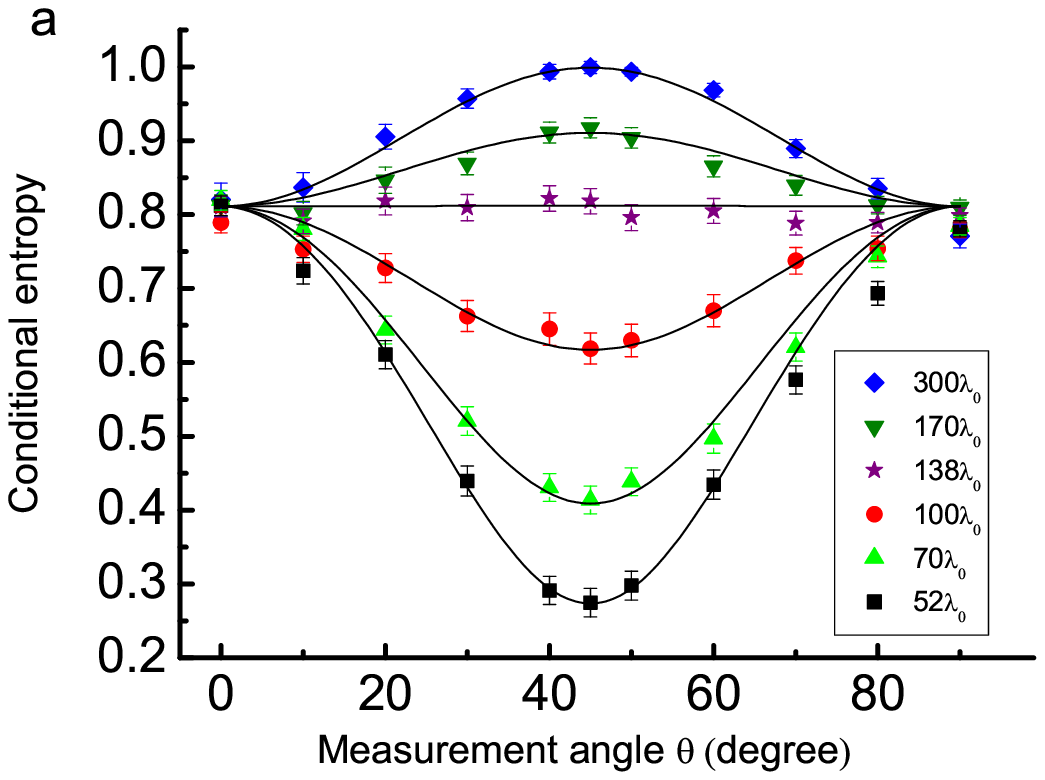}
\end{minipage}\\
\begin{minipage}[c]{0.5\textwidth}
\centering
\includegraphics[width=2.5in]{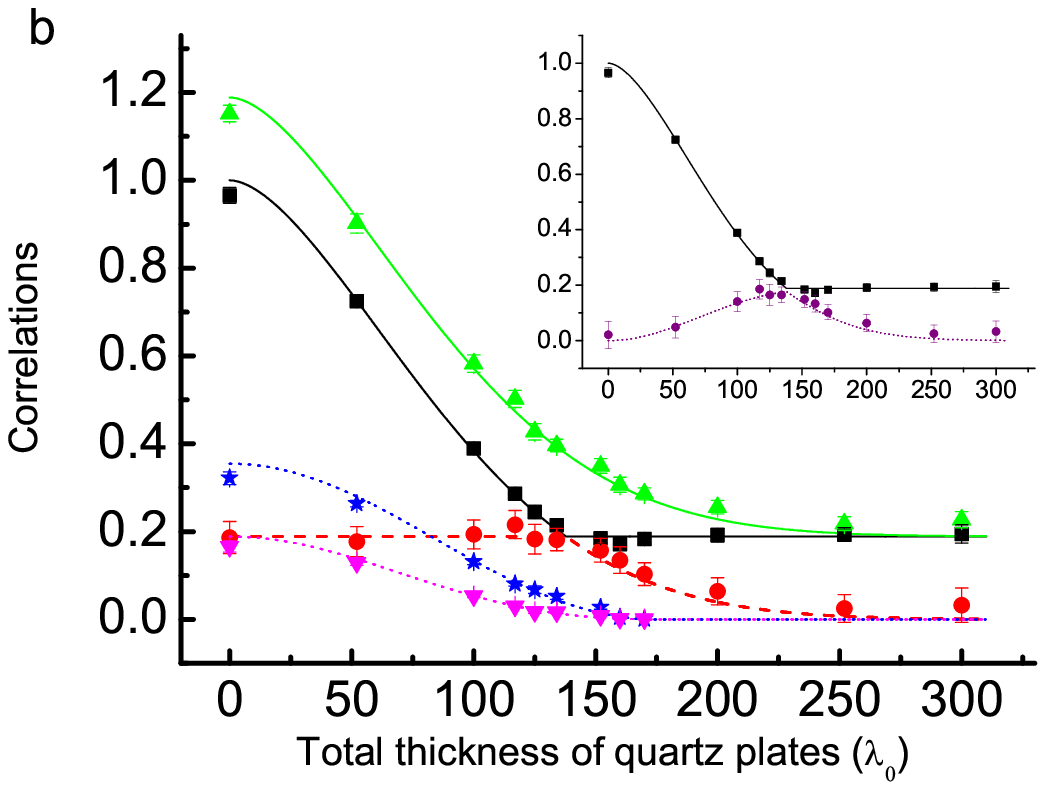}
\end{minipage}
\caption{(Color online). The correlation dynamics of input state
(\ref{state-interference}) with $b=0.75$. ({\bf a}) The values of
conditional entropy $S(\rho_{A}^{l})$, with photons evolving in
different thickness of quartz plates ($L$), as a function of the
measurement angle $\theta$ in mode $B$. ({\bf b}) The dynamics of
correlations. Green upward-pointing triangles, black squares, red
dots, blue stars and magenta downward-pointing triangles represent
experimental results of $\mathcal{I}$, $\mathcal{C}$, $\mathcal{Q}$,
$En$ and $Rn$ with the green solid line, black solid line, red
dashed line, blue dotted line and magenta dotted line representing
the corresponding theoretical predictions. Non-entanglement quantum
correlation ($\mathcal{D}$) is further compared with $\mathcal{C}$
in the inset (the $x$ axis represents the total thickness of quartz
plates and the $y$ axis denotes different kinds of correlations).
Purple dots represent the experimental results of $\mathcal{D}$ and
the purple dotted line is the corresponding theoretical prediction.
Error bars correspond to counting statistics. $\lambda_{0}=0.78$
$\mu$m.} \label{a=0.75}
\end{figure}

Figure 2 shows the correlation dynamics of input state
(\ref{state-interference}) with $b=0.75$ in the phase-damping
channel. Theoretically, the minimum of conditional entropy
$S(\rho_{A}^{l})$ in mode $A$ is obtained by projecting the photon
in mode $B$ onto $|l\rangle=\cos\theta|H\rangle+\sin\theta
e^{i\phi}|V\rangle$, with optimization over the angles $\theta$ and
$\phi$. In our experiment, we measured $S(\rho_{A}^{l})$ as a
function of $\theta$ with $\phi=0$ in different thickness of quartz
plates, as shown in Fig. 2a. We found that the minimum value of
$S(\rho_{A}^{l})$ is obtained with $\theta=45^\circ$ at about
$L<138\lambda_{0}$ and $\theta=0^\circ$ at about
$L\geq138\lambda_{0}$ ($\lambda_{0}=0.78$ $\mu$m is the central
wavelength of the photon), which agree well with theoretical
predictions (solid lines) \cite{Maziero09}. This implies that there
is a sudden change in behavior in the decay rate of $\mathcal{C}$
according to equation (\ref{classical}). The dynamics of total
correlation ($\mathcal{I}$), classical correlation ($\mathcal{C}$)
and quantum correlation ($\mathcal{Q}$) are shown in Fig. 2b. We
find that $\mathcal{C}$ (black squares) decay monotonically at
$L<138\lambda_{0}$, and then remains constant at
$L\geq138\lambda_{0}$, which agrees with Fig. 2a. Such immunity of
classical correlation against decoherence implies an operational way
to compute classical and quantum correlations \cite{Maziero09}. In
contrast, $\mathcal{Q}$ (red dots) behaves in the opposite manner;
it remains constant at $L<138\lambda_{0}$ and decays monotonically
at $L\geq138\lambda_{0}$. $\mathcal{C}$ and $\mathcal{Q}$ overlap at
the specific thickness of about $L=138\lambda_{0}$. It is
interesting to observe the ``decoherence-free" area of quantum
correlation, which is due to the equal decay rate of $\mathcal{I}$
and $\mathcal{C}$. This phenomenon is later shown explicitly in
theory by Mazzola {\it et al.} \cite{Mazzola10}. It may have
important applications for those quantum information protocols using
only quantum correlation. We can see that $\mathcal{I}$ (green
upward-pointing triangles) decays exponentially all the time, which
is consistent with the continuous dynamics of quantum mutual
information according to equation (\ref{total}). The green solid
line, black solid line and red dashed line are the corresponding
theoretical predictions of the total, classical and quantum
correlations. Error bars correspond to counting statistics.

The dynamics of entanglement is also shown in Fig. 2b, which is both
characterized by the entanglement of formation ($En$)
\cite{Bennett96} and the relative entropy of entanglement ($Rn$)
\cite{Vedral97} (the analytical expressions of both
characterizations in our case are given in Methods). The blue stars
in Fig. 2b are the experimental results of $En$ and the blue dotted
line is the corresponding theoretical prediction. The value of
entanglement is set to 0 at a thickness of about $L=173\lambda_{0}$,
which shows the phenomenon of entanglement sudden death \cite{Yu09},
whereas the magenta downward-pointing triangles are the experimental
results of $Rn$. Because the final state in our experiment can be
transformed into a Bell-diagonal form, with a local unitary
operation that does not change its entanglement, the analytical
expressions of $Rn$ (see Methods) are used to give the theoretical
prediction, which is represented by the magenta dotted line.
Although the $En$ is larger than the relative $Rn$ at the beginning
of evolution, they suffer from sudden death at the same thickness,
thereby confirming their self-consistence. We can see that quantum
correlations can be smaller or larger than the $En$ during
evolution. Specifically, quantum correlation decays exponentially,
whereas entanglement disappears completely at the thickness of
$L>173\lambda_{0}$. This confirms the early prediction that quantum
discord is more robust against decoherence than entanglement
\cite{Werlang09,Ferraro09}. Because the relative $Rn$ is on the
equal footing of other measures of correlations in the form of
entropy \cite{Vedral092}, it is not always larger than quantum
correlations, as verified by our experimental results. The inset in
Fig. 2b further compares non-entanglement quantum correlation
defined as $\mathcal{D}=\mathcal{Q}-Rn$ \cite{Vedral092} with
classical correlation. Purple dots are the experimental results of
$\mathcal{D}$, with the purple dotted line representing theoretical
prediction. Because of the sudden change in behaviour of
$\mathcal{Q}$, the decay rate of $\mathcal{D}$ also suffers from
sudden changes and we find that $\mathcal{D}<\mathcal{C}$.

\begin{figure}[tbph]
\begin{center}
\includegraphics [width= 3.0in]{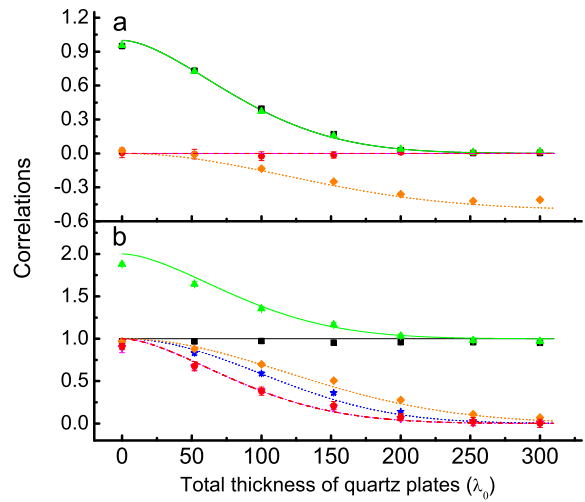}
\end{center}
\caption{(Color online). The correlation dynamics of input state
(\ref{state-interference}) with $b=0.5$ (top panel) and $b=1$
(bottom panel). Green upward-pointing triangles, black squares and
red dots represent the experimental results of $\mathcal{I}$,
$\mathcal{C}$ and $\mathcal{Q}$, respectively. The green solid line,
black solid line and red dashed line are the corresponding
theoretical predictions. $En$ (blue stars) is shown for comparison,
with the blue dotted line representing the corresponding theoretical
prediction. Orange diamonds represent the experimental results of
$\Lambda$ and the orange dotted line is the theoretical prediction.
$\Lambda<0$ in {\bf a} and $En$ is set to 0 (blue stars are not
shown). $Rn$ (magenta downward-pointing triangles) is also shown in
{\bf b} with the magenta dotted line representing the corresponding
theoretical prediction, which completely overlaps with quantum
correlation. Error bars correspond to counting statistics.}
\label{a=0.5}
\end{figure}

Another kind of correlation dynamics is shown in Fig. 3, in which
the sudden change behaviour disappears. The input state is obtained
by equation (\ref{state-interference}) with $b=0.5$ (the separated
state, Fig. 3, top panel) and $b=1$ (the maximally entangled state,
Fig. 3, bottom panel). The minimum of $S(\rho_{A}^{l})$ is gotten
with $\theta=45^\circ$ in Fig. 3 (top panel) and $\theta=0^\circ$ in
Fig. 3 (bottom panel) ($\phi=0$) \cite{Maziero09}. The Green
upward-pointing triangles, black squares and red dots represent the
experimental results of $\mathcal{I}$, $\mathcal{C}$ and
$\mathcal{Q}$, respectively. The quantum correlation in Fig. 3 (top
panel) remains at 0 and the total correlation is equal to the
classical correlation, both of which decay monotonically. Figure 3
(bottom panel) shows another case in which classical correlation
remains at 1 and quantum correlation decays exponentially. Compared
with the dynamics of entanglement $En$ (blue stars), we find that
entanglement also remains at 0, as shown in Fig. 3 (top panel)
(experimental results not shown), which is equal to $\mathcal{Q}$
there, whereas it is larger than $\mathcal{Q}$ during evolution, as
shown in Fig. 3 (bottom panel). Orange diamonds represent the value
of $\Lambda$ (if $\Lambda\geq0$, it represents the quantity of
concurrence \cite{Wootters98}; see Methods), which also decays
exponentially. The value of $En$ is set to 0 when $\Lambda<0$
according to its definition. The relative $Rn$ (magenta
downward-pointing triangles) is also shown in Fig. 3. It remains
constant at 0, as shown in Fig. 3 (top panel) (experimental results
not shown) and then decays exponentially, as shown in Fig. 3 (bottom
panel). $Rn$ completely overlaps with quantum correlation in both
cases, and the non-entanglement quantum correlation reads at 0. The
experimental results agree well with the corresponding theoretical
predictions.

\begin{figure}[tbph]
\begin{center}
\includegraphics [width= 3.0in]{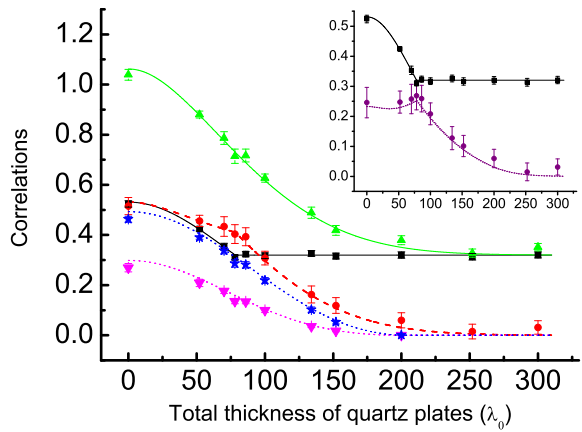}
\end{center}
\caption{(Color online). The correlation dynamics of input state
(\ref{four-final}) with $b=0.9$ and $R=0.9$. Green upward-pointing
triangles, black squares, red dots, blue stars and magenta
downward-pointing triangles represent the experimental values of
$\mathcal{I}$, $\mathcal{C}$, $\mathcal{Q}$, $En$ and $Rn$,
respectively. The green solid line, black solid line, red dashed
line, blue dotted line and magenta dotted line are the corresponding
theoretical predictions. The inset (the $x$ axis represents the
total thickness of quartz plates and the $y$ axis denotes different
kinds of correlations) shows the comparison of $\mathcal{D}$ (purple
dots) with $\mathcal{C}$. The purple dotted line is the theoretical
prediction of $\mathcal{D}$. Error bars correspond to counting
statistics.} \label{a=0.9}
\end{figure}

We further show an interesting case in which $\mathcal{Q}$ can be
larger than $\mathcal{C}$. In this case, the dotted pane is inserted
into mode $B$, as shown in Fig. \ref{fig:setup} and the initial
input state is given by equation (\ref{four-final}). We set $b=0.9$
and $R=0.9$ to maximize the value of $\mathcal{Q}-\mathcal{C}$. Fig.
4 shows our experimental results. The total correlation (green
upward-pointing triangles) decays exponentially, whereas the
classical (black squares) and quantum (red dots) correlations
exhibit sudden changes in behaviour in their decay rates at about
$L=78\lambda_{0}$. After this special thickness, the classical
correlation remains constant and the quantum correlation decays
exponentially with a higher decay rate. During the evolution,
$\mathcal{Q}$ is observed to be larger than $\mathcal{C}$ at the
thickness interval from about $50\lambda_{0}$ to $90\lambda_{0}$,
within error bars. Such an observation contradicts the early
conjecture that $\mathcal{C}\geq\mathcal{Q}$ \cite{Lindblad91} and
it is consistent with previous theoretical predictions
\cite{Sarandy09,Maziero09,Luo09}. The dynamics of entanglement,
which are characterized both by $En$ (blue stars) and $Rn$ (magenta
down triangles), are also shown in Fig. 4. Both characterizations
suffer from sudden death at the same thickness of about
$L=202\lambda_{0}$ and the entanglement is set to 0 in the
subsequent evolution. The value of the non-entanglement quantum
correlations $\mathcal{D}$ (purple dots), which also exhibit a
sudden change in the decay rate, is compared with the classical
correlation in the inset. We find that entanglement and
non-entanglement quantum correlations are always less than the
classical correlation in our experiment.

\section{Discussion}

The classical and quantum correlations considered in this experiment
are defined by the one-partition measurement of a bipartite system
\cite{Henderson01,Ollivier01}. In such cases, analytical solutions
for classical and quantum correlations have been obtained for some
kinds of states with a high symmetry
\cite{Luo082,Maziero09,Sarandy09}. Although classical and quantum
correlations may be generally asymmetric according to the choice of
partition to be measured, for states with maximally mixed marginals,
they are symmetric under the interchange of the measured side
\cite{Maziero092}. There is also a quantifier of classical
correlation with measurement over both partitions of a bipartite
system, which is defined as the maximum classical mutual information
\cite{Terhal02,DiVincenzo04}. For states with maximally mixed
marginals, the definitions of classical and quantum correlations are
numerically verified to be equal to that defined with one-side
measurement \cite{Maziero092}. A thermodynamic approach is also
developed to quantify correlations \cite{Oppenheim02,Horodecki05}.
In particular, the quantum-information deficit is defined to
characterize the difference between the information that can be
localized with Closed Local Operations and Classical Communications
and the total information of the state, and it quantifies the
quantumness of correlations \cite{Horodecki05}. We find that the
quantum-information deficit is equal to quantum discord for
Bell-diagonal states. On the basis of the idea that classical states
can be measured without disturbance, Luo \cite{Luo08} use the
measurement-induced disturbance to characterize classical and
quantum correlations. Because of the symmetric property of
Bell-diagonal states, these definitions of correlations used in our
experiment also coincide with that defined by Luo's method
\cite{Modi10}. Recently, a method to quantify quantum and classical
correlations on an equal footing with relative entropy has been
proposed \cite{Modi10}, and it is applicable to multi-particle
systems of arbitrary dimensions. In our case, considering the
dynamics of Bell-diagonal states, the one-side definitions of
classical and quantum correlations \cite{Henderson01,Ollivier01} are
consistent with those defined by the equal footing method
\cite{Modi10}. As a result, for the Bell-diagonal states used in our
experiment, all of the above separation methods for classical and
quantum correlations are coincident.

Through measuring classical and quantum correlations in multipartite
systems, Bennett {\it et al.} \cite{Bennett10} recently introduced
three postulates that should be satisfied by any measure or
indicator of genuine multipartite correlations. They found that the
concept of covariance proposed by Kaszlikowski {\it et al.}
\cite{Kaszlikowski08} to identify the existence of genuine
multipartite correlations does not satisfy two postulates, and they
concluded that it cannot be an indicator of the genuine multipartite
classical correlations \cite{Bennett10}. Because the equal footing
method mentioned above can be directly applied to multipartite
systems with a unified view of quantum and classical correlations
\cite{Modi10}, it may inspire insight into current debates.

The relationship between the $En$ and quantum correlation is also
interesting. In our experiment, the $En$ is observed to be even
larger than quantum correlation in certain areas of evolution. For
bipartite systems, it has also been shown that quantum mutual
information considered as the measurement of total correlation can
be smaller than the $En$ \cite{Hayden06,Li07}, which may occur in
the case of particles with dimensions larger than five \cite{Li07}.
Such a contradiction is due to the different frameworks of their
definitions. In our case, we can choose the relative $Rn$ to measure
entanglement, so as to make it on an equal footing with other
measures of correlations in the form of entropy \cite{Vedral092}.

In conclusion, we have shown the sudden change in behaviour in the
decay rates of different kinds of correlations in the two-photon
system. Our results show that classical and quantum correlations can
remain unaffected under certain decoherence areas, which will play
important roles in distinguishing correlations \cite{Maziero09} and
designing robust quantum protocols based only on quantum
correlation. The early conjecture that $\mathcal{C}\geq\mathcal{Q}$
\cite{Lindblad91} is disproved by our experimental results. On the
other hand, classical correlation is always found to be larger than
the entanglement and non-entanglement quantum correlation in the
experiment.

\section{Methods}

{\bf Analytical expressions of the $En$ and the relative $Rn$.} For
biqubit systems, the $En$ can be given by the analytical formula
\cite{Wootters98}
\begin{equation}
En(\rho)=H(\frac{1+\sqrt{1-\Upsilon^2}}{2}) \label{Eof}
\end{equation}
where $H(x)=-x\log_{2}x-(1-x)\log_{2}(1-x)$. $\Upsilon$ is the
concurrence, given by $\Upsilon=\textrm{max}\{0,\Lambda\}$, where
$\Lambda=\sqrt{\chi_{1}}-\sqrt{\chi_{2}}-\sqrt{\chi_{3}}-\sqrt{\chi_{4}}$
and $\chi_{j}$ are the eigenvalues in decreasing order of the matrix
$\rho(\sigma_{2}\otimes\sigma_{2})\rho^{\ast}(\sigma_{2}\otimes\sigma_{2})$
with $\sigma_{2}$ denoting the second Pauli matrix and $\rho^{*}$
corresponding to the conjugate of $\rho$ in the canonical basis
$\{|HH\rangle,|HV\rangle,|VH\rangle,|VV\rangle\}$. In addition, the
relative $Rn$ is defined as the minimal distance between $\rho$ and
a disentangled state \cite{Vedral97}. For the Bell-diagonal state,
if the four eigenvalues of the density matrix are
$\lambda_{j=1}^{4}\in[0,1/2]$, the relative $Rn$ is given by
\cite{Vedral97}
\begin{equation}
Rn(\rho)=0, \label{REE1}
\end{equation}
whereas for any $\lambda_{j}\geq1/2$ we obtain
\begin{equation}
Rn(\rho)=\lambda_{j}\log_{2}\lambda_{j}+(1-\lambda_{j})\log_{2}(1-\lambda_{j})+1.
\label{REE2}
\end{equation}

\section{Acknowledgments}

We thank V. Vedral, S. Luo, and Y.-C. Wu for helpful discussions.
This work was supported by National Fundamental Research Program,
National Natural Science Foundation of China (Grant No.60121503,
10874162).

\end{document}